\documentclass{conm-p-l}

\usepackage{amssymb}

\usepackage{accents} 
\usepackage{nicefrac} 

\copyrightinfo{2010}{A. Garrett Lisi}

\theoremstyle{definition}

\theoremstyle{remark}

\numberwithin{equation}{section}

\newcommand{\beq}{\begin{equation}}
\newcommand{\eeq}{\end{equation}}

\def \ga{\gamma}
\def \de{\delta}
\def \ep{\epsilon}

\def \et{\eta}

\def \io{\iota}
\def \ka{\kappa}
\def \la{\lambda}

\def \si{\sigma}

\def \ph{\phi}

\def \ps{\psi}
\def \om{\omega}
\def \Ga{\Gamma}

\def \La{\Lambda}

\def \pa{\partial}

\def \lb{\left[}
\def \rb{\right]}
\def \lp{\left(}
\def \rp{\right)}

\def \p#1{\phantom{#1}}
\def \vp#1{\vphantom{#1}}
\def \fr#1#2{{\textstyle \frac{#1}{#2}}}
\def \ha{\fr{1}{2}}

\def \f#1{\underaccent{\!\p{.}\! \raisebox{0pt}{\_} }{{#1}}}
\def \ff#1{\underaccent{=}{{#1}}}

\def \ve#1{\accentset{\rightharpoonup}{{#1}}}

\def \ud#1{\underaccent{.}{{#1}}}
\def \udf#1{\underaccent{\_ .}{{#1}}}
\def \udff#1{\underaccent{= \!\p{.}\! \smash{\cdot}}{{#1}}}

\begin{document}

\title{An Explicit Embedding of Gravity and the Standard Model in E8}


\author{A. Garrett Lisi}
\address{A. Garrett Lisi, 95 Miner Pl, Makawao, HI 96768}
\email{alisi@hawaii.edu}


\date{}

\begin{abstract}
The algebraic elements of gravitational and Standard Model gauge fields acting on a generation of fermions may be represented using real matrices. These elements match a subalgebra of spin(11,3) acting on a Majorana-Weyl spinor, consistent with GraviGUT unification. This entire structure embeds in the quaternionic real form of the largest exceptional Lie algebra, E8. These embeddings are presented explicitly and their implications discussed.
\end{abstract}

\maketitle


\section{Introduction}

Since the inception of the Standard Model of particle physics in the 1970s, theorists have speculated that the zoo of elementary particles might correspond to a more unified mathematical structure. The most well established of these unification attempts are the classical Grand Unified Theories (GUTs), involving the embedding of the Standard Model Lie algebra, $su(3) \!\oplus\! su(2) \!\oplus\! u(1)$, in larger Lie algebras.\cite{Baez}  Some evidence that this unification occurs in nature comes from the fact that the coupling strengths of the strong $su(3)$ and electroweak $su(2) \!\oplus\! u(1)$ forces converge to approximately the same value for short distances. Also, the fermions, which live in representation spaces of the Standard Model Lie group, embed nicely in representation spaces of the unified groups.

In the Georgi-Glashow Grand Unified Theory, the Standard Model Lie algebra embeds in $su(5)$ and the fermions live in $\bar{5}$ and $10$ representation spaces. Unfortunately for this GUT, the new particles in $su(5)$ would allow protons to decay at a rapid rate, which has been ruled out by experiment. In another Grand Unified Theory, which has not yet been ruled out by proton decay, the Standard Model Lie algebra embeds in $spin(10)$ and fermions live in a $16$ spinor rep. This $spin(10)$ GUT contains the $su(5)$ GUT as a subalgebra and also contains a third GUT, the Pati-Salam GUT, via
$$
su(4) \!\oplus\! su(2) \!\oplus\! su(2) = spin(6) \!\oplus\! spin(4) \subset spin(10)
$$
All of these Grand Unified Theories are good models for unifying the strong and electroweak forces, but disregard gravity.

Geometrically, the gauge fields of the Standard Model are parts of the connection of a principal $SU(3)\!\otimes\!SU(2)\!\otimes\!U(1)$ bundle over four dimensional spacetime. Algebraically, this gauge connection is a 1-form valued in the Standard Model Lie algebra. In modern descriptions of gravity the gravitational field is also a connection -- the {\it spin connection} 1-form, valued in the $spin(1,3)$ Lie algebra -- which acts on Dirac fermions in 4-spinor reps.  These spin and gauge connections correspond to the complete structure group, $SU(3)\!\otimes\!SU(2)\!\otimes\!U(1)\!\otimes\!Spin(1,3)$ (modulo discrete subgroups), of the associated fermion vector bundle. This structure group also acts on the {\it gravitational frame} -- a 1-form field valued in the 4-vector of $spin(1,3)$ -- and on a set of Higgs scalar fields.

Given the success of the $spin(10)$ Grand Unified Theory, it is natural to consider its further unification with gravity in a $spin(11,3)$ GraviGUT, with each generation of fermions in a 64-spinor rep.\cite{Percacci} In such a GraviGUT, the gravitational frame and Higgs fields naturally reside in the complement of $spin(1,3)$ and $spin(10)$ in $spin(11,3)$, as the {\it frame-Higgs} part of the {\it unified bosonic connection}.\cite{BF, LSS} Geometrically, all elementary particle fields correspond to parts of the unified bosonic connection or the fermion fibers over spacetime. Algebraically, elementary particles are eigenstates (weight vectors) of the gravitational and Standard Model Lie algebras, and their eigenvalues (weights) are their charges with respect to the different forces.

In previous work it was proposed that all bosons {\it and} fermions may be unified as parts of a {\it superconnection} valued in the largest simple exceptional Lie algebra, $E_8$.\cite{E8} This proposal was based on the observation that the weights of gravitational and Standard Model bosons and fermions match roots of $E_8$. Such a description, using weight matching, is a valuable initial step, but does not fully account for the non-compact nature of the groups involved.

In this new work, a companion to \cite{E8}, we provide an explicit match between algebraic elements of gravity and the Standard Model and some Lie algebra generators of $E_{8(-24)}$. This match is achieved via the intermediate step of embedding gravity and the Standard Model in the $spin(11,3)$ GraviGUT using explicit matrix representations. Throughout the exposition, terminology is purposefully kept as elementary, complete, and explicit as possible, in the interest of accessibility.

\section{Algebra of Gravity and the Standard Model}

The matrix representation building blocks of spin algebras are the traceless, Hermitian, $\mathbb{C}(2)$ Pauli matrices,
$$
\si_1 = \lb \begin{matrix} 0 & 1 \cr 1 & 0 \cr \end{matrix} \rb
\;\;\;\;\;\;\;\;\;
\si_2 = \lb \begin{matrix} 0 & -i \cr i & 0 \cr \end{matrix} \rb
\;\;\;\;\;\;\;\;\;
\si_3 = \lb \begin{matrix} 1 & 0 \cr 0 & -1 \cr \end{matrix} \rb
$$
which, under the matrix product, satisfy
\beq
\si_I \si_J = \de_{IJ} 1 + i \, \ep_{IJ}^{\p{IJ}K} \si_K
\label{pid}
\eeq
in which ``$1$'' stands for an identity matrix of appropriate size, and $\ep^{IJK}$ is the permutation symbol. These three $\mathbb{C}(2)$ matrices, $\si_I$, represent basis vector elements of the $Cl(3)$ Clifford algebra.

\subsection{Gravity}

Using the tensor product of Pauli matrices, the four basis vectors of the spacetime Clifford algebra, $Cl(1,3)$, are conventionally represented in $\mathbb{C}(4)$ as
\beq
\begin{array}{rcrccc}
\ga_1 \!\!&\!\!=\!\!&\!\!  i \!\!\!&\!\! \si_2 \!\!\!&\!\! \otimes \!\!\!&\!\! \si_1 \cr
\ga_2 \!\!&\!\!=\!\!&\!\!  i \!\!\!&\!\! \si_2 \!\!\!&\!\! \otimes \!\!\!&\!\! \si_2 \cr
\ga_3 \!\!&\!\!=\!\!&\!\!  i \!\!\!&\!\! \si_2 \!\!\!&\!\! \otimes \!\!\!&\!\! \si_3 \cr
\ga_4 \!\!&\!\!=\!\!&\!\!    \!\!\!&\!\! \si_1 \!\!\!&\!\! \otimes \!\!\!&\!\! 1 \cr
\end{array}
\;\;\;\;
v^a \ga_a =
\lb \begin{matrix}
0 & 0 & v^3 + v^4 & v^1 - i \, v^2 \cr
0 & 0 & v^1 + i \, v^2 & -v^3 + v^4 \cr
-v^3 + v^4 & -v^1 + i \, v^2 & 0 & 0 \cr
-v^1 -i \, v^2 & v^3 + v^4 & 0 & 0 \cr
\end{matrix} \rb 
\label{cl13}
\eeq
the chiral Weyl representation of the Dirac matrices. Taking the product of these, using (\ref{pid}) on both levels, the $\mathbb{C}(4)$ matrix representatives of the six independent $Cl^2(1,3) = spin(1,3)$ bivector basis elements
$$
\ga_{ab} = \ga_{[ a} \ga_{b ]} = \ha ( \ga_a \ga_b - \ga_b \ga_a ) \;\;\;\;\; = \ga_a \ga_b \;\; \mbox{for} \; a < b
$$
are
$$
\begin{array}{rcrccc}
\ga_{23} \!\!&\!\!=\!\!&\!\! -i \!\!\!&\!\! 1 \!\!\!&\!\! \otimes \!\!\!&\!\! \si_1 \cr
\ga_{13} \!\!&\!\!=\!\!&\!\!  i \!\!\!&\!\! 1 \!\!\!&\!\! \otimes \!\!\!&\!\! \si_2 \cr
\ga_{12} \!\!&\!\!=\!\!&\!\! -i \!\!\!&\!\! 1 \!\!\!&\!\! \otimes \!\!\!&\!\! \si_3 \cr
\ga_{14} \!\!&\!\!=\!\!&\!\!    \!\!\!&\!\! \si_3 \!\!\!&\!\! \otimes \!\!\!&\!\! \si_1 \cr
\ga_{24} \!\!&\!\!=\!\!&\!\!    \!\!\!&\!\! \si_3 \!\!\!&\!\! \otimes \!\!\!&\!\! \si_2 \cr
\ga_{34} \!\!&\!\!=\!\!&\!\!    \!\!\!&\!\! \si_3 \!\!\!&\!\! \otimes \!\!\!&\!\! \si_3 \cr
\end{array}
$$
With these basis elements, the gravitational spin connection, written locally as $\f{\om} = \ha \f{dx^\mu} \om_\mu^{\p{\mu}ab} \ga_{ab}$, acts on fermions as $4^\mathbb{C}_S$ Dirac spinors,
$$
\f{\om} \, \ps =
\lb \begin{matrix}
\f{\om}{}_L & 0 \cr
 0 & \f{\om}{}_R \cr
\end{matrix} \rb
\lb \begin{matrix}
\ps_L \cr
\ps_R \cr
\end{matrix} \rb 
$$
in which the left (positive) chiral $2^\mathbb{C}_{S+}$ and right (negative) chiral $2^\mathbb{C}_{S-}$ parts of the Dirac spinor are acted on independently as
\beq
\f{\om}{}_L \ps_L =
\lb \begin{matrix}
-i \, \f{\om}^{12}+\f{\om}^{34} & \f{\om}^{13}-i\,\f{\om}^{23}+\f{\om}^{14}-i\,\f{\om}^{24} \cr
-\f{\om}^{13}-i\,\f{\om}^{23}+\f{\om}^{14}+i\,\f{\om}^{24} & i\,\f{\om}^{12}-\f{\om}^{34} \cr
\end{matrix} \rb
\lb \begin{matrix}
\ps_L^\wedge \cr
\ps_L^\vee \cr
\end{matrix} \rb 
\label{omL}
\eeq
and 
$$
\f{\om}{}_R \ps_R =
\lb \begin{matrix}
-i \, \f{\om}^{12}-\f{\om}^{34} & \f{\om}^{13}-i\,\f{\om}^{23}-\f{\om}^{14}+i\,\f{\om}^{24} \cr
-\f{\om}^{13}-i\,\f{\om}^{23}-\f{\om}^{14}-i\,\f{\om}^{24} & i\,\f{\om}^{12}+\f{\om}^{34} \cr
\end{matrix} \rb
\lb \begin{matrix}
\ps_R^\wedge \cr
\ps_R^\vee \cr
\end{matrix} \rb 
$$
with $\ps^\wedge$ and $\ps^\vee$ the ``spin up'' and ``spin down'' components of a fermion. Note that, with our choice of representation, $\f{\om}{}_R = - \f{\om}{}_L^{\dagger}$; and the traceless $\mathbb{C}(2)$ representation in (\ref{omL}) establishes the $spin(1,3) = sl(2,\mathbb{C})$ Lie algebra equivalence.

The $\mathbb{C}(4)$ chirality projection operators,
\beq
P_{L/R} = \ha (1 \pm i \, \ga)
\label{PL}
\;\;\;\;\;\;\;\;
P_L =
\lb \begin{matrix}
1 & 0 \cr
0 & 0 \cr
\end{matrix} \rb
\;\;\;\;\;\;\;\;
P_R =
\lb \begin{matrix}
0 & 0 \cr
0 & 1 \cr
\end{matrix} \rb
\eeq
are constructed using the $Cl^4(1,3)$ pseudoscalar,
$$
\ga = \ga_1 \ga_2 \ga_3 \ga_4 = -i \, \si_3 \otimes 1 =
\lb \begin{matrix}
-i & 0 \cr
0 & i \cr
\end{matrix} \rb
$$
and are needed to describe the action of Standard Model gauge fields on fermions.

\subsection{The Standard Model}

The three conventional basis $\mathbb{C}(2)$ matrix generators of the real $su(2)$ Lie algebra are $\fr{i}{2} \si_I$, with their Lie bracket satisfying
$$
[ \fr{i}{2} \si_I, \fr{i}{2} \si_J ] = - \fr{1}{4} ( \si_I \si_J - \si_J \si_I )= - \ep_{IJ}^{\p{IJ}K} \fr{i}{2} \si_K
$$
with real structure constants, $- \ep_{IJ}^{\p{IJ}K}$. Curiously, in the Standard Model, the electroweak $su(2)$ acts only on the left chiral parts of leptons and quarks arranged in doublets, such as on $\lb \begin{matrix} \nu_L \cr e_L \cr \end{matrix} \rb$ and $\lb \begin{matrix} u_L \cr d_L \cr \end{matrix} \rb$. Acting on a doublet $8^\mathbb{C}$ column of two Dirac fermions, such as $\lb \begin{matrix} \nu \cr e \cr \end{matrix} \rb$, the electroweak $su(2)_L$ generators are $\fr{i}{2} \si_I \!\otimes\! P_L$, employing the left chiral projection matrix (\ref{PL}).

The real $su(3)$ Lie algebra corresponding to the strong force of the Standard Model acts in the fundamental 3 representation on ``colored'' quarks. The eight $su(3)$ basis generators, $\fr{i}{2} \la_A$, are expressed in terms of the conventional traceless, Hermitian, $\mathbb{C}(3)$ Gell-Mann matrices, $\la_A$, written here in condensed form as
$$
g^A \la_A =
\lb
\begin{array}{ccc}
 g^3 \!+\! \fr{1}{\sqrt{3}} g^8 \! & g^1\!-\!ig^2 & g^4\!-\!ig^5 \\
g^1\!+\!ig^2 & \! -g^3 \!+\! \fr{1}{\sqrt{3}} g^8 & g^6\!-\!ig^7 \\
g^4\!+\!ig^5 & g^6\!+\!ig^7 & \fr{-2}{\sqrt{3}} g^8
\end{array}
\rb
$$
When considering how $su(3)$ acts on all fermions, it is convenient to define the $\mathbb{C}(4)$, ``expanded'' Gell-Mann matrices,
$$
\La_A =
\lb
\begin{array}{cc}
0 & 0 \cr
0 & \la_A \cr
\end{array}
\rb
$$
which act on a multiplet of leptons and red, green, and blue quarks in a $4^\mathbb{C}$ column,
$$
[
\begin{array}{cccc}
l & q^r & q^g & q^b \cr
\end{array}
]
$$
(written here in a row for convenience). These eight expanded Gell-Mann matrix generators span an $su(3)$ subalgebra of $su(4)$.

The electroweak $u(1)$ of the Standard Model acts on fermions according to their weak hypercharge, $Y$. With one complete generation of Standard Model fermions, including left and right chiral components, in a $16^\mathbb{C}$ column,
\beq
\;\;\;\;\;\;\;\;\;
[
\begin{array}{cccccccccccccccc}
\!\! \nu_L \!\!&\!\! \nu_R \!\!&\!\! e_L \!\!&\!\! e_R
\!\!&\!\! u_L^r \!\!&\!\! u_R^r \!\!&\!\! d_L^r \!\!&\!\! d_R^r
\!\!&\!\! u_L^g \!\!&\!\! u_R^g \!\!&\!\! d_L^g \!\!&\!\! d_R^g
\!\!&\!\! u_L^b \!\!&\!\! u_R^b \!\!&\!\! d_L^b \!\!&\!\! d_R^b \!\! \cr
\end{array}
]
\label{ferms}
\eeq
the electroweak $u(1)_Y$ generator acts on this as the diagonal matrix of their hypercharges,
\beq
\fr{i}{2} \, \mbox{diag}(
\begin{array}{cccccccccccccccc}
\!\! \!-1 \!\!&\!\! \,\,0\, \!\!&\!\! \!-1 \!\!&\!\! \!-2
\!\!&\!\! \;\fr{1}{3}\,\, \!\!&\!\! \;\fr{4}{3}\,\, \!\!&\!\! \;\fr{1}{3} \!\!&\!\! -\fr{2}{3}
\!\!&\!\! \;\fr{1}{3}\,\, \!\!&\!\! \;\fr{4}{3}\,\, \!\!&\!\! \;\fr{1}{3} \!\!&\!\! -\fr{2}{3}
\!\!&\!\! \;\fr{1}{3}\,\, \!\!&\!\! \;\fr{4}{3}\,\, \!\!&\!\! \;\fr{1}{3} \!\!&\!\! -\fr{2}{3}
\!\!
\end{array}
) 
\label{Y}
\eeq
This $\mathbb{C}(16)$ generator (or a $\mathbb{C}(32)$ generator if spin components are included), with these hypercharges, happens to be the combination of an $su(2)_R$ generator and an $su(4)$ generator,
$$
\fr{i}{2} ( 1 \!\otimes\! \si_3 \!\otimes\! P_R + \sqrt{\fr{2}{3}} \, \La_{15} \!\otimes\! 1 \!\otimes\! 1 )
$$
in which
$$
\La_{15} = \sqrt{\fr{3}{2}} \, \mbox{diag}(
\begin{array}{cccc}
\!\! -1\,\, \!\!&\!\! \;\fr{1}{3}\,\, \!\!&\!\! \;\fr{1}{3}\,\, \!\!&\!\! \;\fr{1}{3}\,\, \!\!
\end{array}
)
$$
is an ``extended'' Gell-Mann matrix, related to the $\fr{i}{2} \La_{15}$ generator of $su(4)$. This $\fr{i}{2} \La_{15}$ generator of $su(4)$ corresponds to a $u(1)_{B-L}$ subalgebra, with charges corresponding to baryon minus lepton number. This peculiar coincidence of hypercharges, (\ref{Y}), is the key to the successful embedding of the Standard Model Lie algebra in Grand Unified Theories.

The familiar electric charges of elementary particles are the components of the $u(1)_Q$ generator, a combination of the $u(1)_Y$ generator, (\ref{Y}), and an $su(2)_L$ generator,
$$
\begin{array}{cl}
\!\!&\!\! \fr{i}{2} ( 1 \!\otimes\! \si_3 \!\otimes\! P_R + \sqrt{\fr{2}{3}} \, \La_{15} \!\otimes\! 1 \!\otimes\! 1 )
+ \fr{i}{2} \, 1 \!\otimes\! \si_3 \!\otimes\! P_L \vp{a_{\Big(}} \cr
= \!\!&\!\! i \, \mbox{diag}(
\begin{array}{cccccccccccccccc}
\!\! 0\, \!\!&\!\! \,\,0\, \!\!&\!\! \!-1 \!\!&\!\! \!-1
\!\!&\!\! \;\fr{2}{3}\, \!\!&\!\! \;\fr{2}{3}\, \!\!&\!\! -\fr{1}{3} \!\!&\!\! -\fr{1}{3}
\!\!&\!\! \;\fr{2}{3}\, \!\!&\!\! \;\fr{2}{3}\, \!\!&\!\! -\fr{1}{3} \!\!&\!\! -\fr{1}{3}
\!\!&\!\! \;\fr{2}{3}\, \!\!&\!\! \;\fr{2}{3}\, \!\!&\!\! -\fr{1}{3} \!\!&\!\! -\fr{1}{3}
\!\!
\end{array}
)
\end{array}
$$

\subsection{Together}

Each generation of Standard Model fermions is a $32^\mathbb{C}$ column, with the field degrees of freedom of the first generation fermions labeled according to (\ref{ferms}), with two spin components included for each.  The six gravitational $spin(1,3)$, eight strong $su(3)$, three electroweak $su(2)_L$, and one hypercharge $u(1)_Y$ basis generators of the combined $su(3) \!\oplus\! su(2)_L \!\oplus\! u(1)_Y \!\oplus\! spin(1,3)$ Lie algebra of gravity and the Standard Model act on these as eighteen $\mathbb{C}(32)$ matrix generators:
\beq
\begin{array}{rcrcccccl}
T_{ab}^G \!\!&\!\! = \!\!&\!\! \!\!&\!\! 1 \!\!&\!\! \!\otimes\! \!\!&\!\! 1 \!\!&\!\! \!\otimes\! \!\!&\!\! \ga_{ab} \!\!& \vp{a_{\big(}}\cr
T_{A}^g \!\!&\!\! = \!\!&\!\! \fr{i}{2}\! \!\!&\!\! \La_A \!\!&\!\! \!\otimes\! \!\!&\!\! 1 \!\!&\!\! \!\otimes\! \!\!&\!\! 1  \!\!& \vp{a_{\big(}} \cr
T_{I}^W \!\!&\!\! = \!\!&\!\! \fr{i}{2}\! \!\!&\!\! 1 \!\!&\!\! \!\otimes\! \!\!&\!\! \si_I \!\!&\!\! \!\otimes\! \!\!&\!\! P_L  \!\!& \vp{a_{\big(}}\cr
T^Y \!\!&\!\! = \!\!&\!\! \fr{i}{2}\! \!\!&\!\! 1 \!\!&\!\! \!\otimes\! \!\!&\!\! \si_3 \!\!&\!\! \!\otimes\! \!\!&\!\! P_R  \!\!&\!\!
+ \; \fr{i}{2}  \sqrt{\fr{2}{3}} \, \La_{15} \otimes 1 \otimes 1
\end{array}
\label{gens}
\eeq
The spin and gauge parts of the connection, $( \ha \f{\om} + \f{g} + \f{W} + \f{B})$, in terms of these basis generators, are $\f{\om} = \ha \f{\om}^{ab} T_{ab}^G$ for the $spin(1,3)$ connection, $\f{g}=\f{g}^A T_A^g$ for the $su(3)$ gluons, $\f{W}=\f{W}^I T_I^W$ for $su(2)_L$, and $\f{B}=\f{B}^Y T^Y$ for $u(1)_Y$.

Looking at these eighteen generators, as $\mathbb{C}(32)$ matrices, each has either real or imaginary components. And these act on a generation of Standard Model fermions in a $32^\mathbb{C}$ representation space. Nevertheless, they span a real Lie algebra, with real structure constants and a corresponding real Lie group. In order to represent these generators with real matrices, we use a standard trick to ``realize'' the complex structure. We convert every $i$ in the generator matrices (\ref{gens}) to an $\mathbb{R}(2)$ matrix squaring to minus the identity,
$$
i \mapsto
\lb \begin{array}{cc}
0 & -1 \cr
1 & 0 \cr
\end{array} \rb
= -i \si_2
\;\;\;\;\;\;\;\;\;\;\;
f \mapsto
\lb \begin{array}{cc}
f_r \cr
f_i \cr
\end{array} \rb
$$
and convert every complex fermion component to a real $2^\mathbb{R}$, so that
$$
i f = \lb \begin{array}{cc}
0 & -1 \cr
1 & 0 \cr
\end{array} \rb
\lb \begin{array}{cc}
f_r \cr
f_i \cr
\end{array} \rb
=
\lb \begin{array}{cc}
-f_i \cr
f_r \cr
\end{array} \rb
= i (f_r+i f_i) = - f_i + i f_r
$$
With this conversion, the basis generators (\ref{gens}) become $\mathbb{R}(64)$ matrices, corresponding to the same Lie algebra, and act on a generation of fermions in a $64^\mathbb{R}$ in the same way as the $\mathbb{C}(32)$ generators acted on a $32^\mathbb{C}$. To be complete and explicit, the eighteen basis generators (\ref{gens}) of the gravitational and Standard Model Lie algebra, represented as $\mathbb{R}(64)$ matrices, may be expressed numerically using the tensor products (at six levels) of $\mathbb{C}(2)$ Pauli matrices. These matrices, and a $64^\mathbb{R}$ of the first generation fermions, are:
\beq
\renewcommand{\arraystretch}{1.25}
\begin{array}{rcrccccccccccc}
T^G_{23} \!\!&\!\! = \!\!&\!\! i \!\!&\!\! 1 \!\!\!&\!\! \otimes \!\!\!&\!\! 1 \!\!\!&\!\! \otimes \!\!\!&\!\! 1 \!\!\!&\!\! \otimes \!\!\!&\!\! 1 \!\!\!&\!\! \otimes \!\!\!&\!\! \si_1 \!\!\!&\!\! \otimes \!\!\!&\!\! \si_2 \cr
T^G_{13} \!\!&\!\! = \!\!&\!\! i \!\!&\!\! 1 \!\!\!&\!\! \otimes \!\!\!&\!\! 1 \!\!\!&\!\! \otimes \!\!\!&\!\! 1 \!\!\!&\!\! \otimes \!\!\!&\!\! 1 \!\!\!&\!\! \otimes \!\!\!&\!\! \si_2 \!\!\!&\!\! \otimes \!\!\!&\!\! 1 \cr
T^G_{12} \!\!&\!\! = \!\!&\!\! i \!\!&\!\! 1 \!\!\!&\!\! \otimes \!\!\!&\!\! 1 \!\!\!&\!\! \otimes \!\!\!&\!\! 1 \!\!\!&\!\! \otimes \!\!\!&\!\! 1 \!\!\!&\!\! \otimes \!\!\!&\!\! \si_3 \!\!\!&\!\! \otimes \!\!\!&\!\! \si_2 \cr
T^G_{14} \!\!&\!\! = \!\!&\!\!   \!\!&\!\! 1 \!\!\!&\!\! \otimes \!\!\!&\!\! 1 \!\!\!&\!\! \otimes \!\!\!&\!\! 1 \!\!\!&\!\! \otimes \!\!\!&\!\! \si_3 \!\!\!&\!\! \otimes \!\!\!&\!\! \si_1 \!\!\!&\!\! \otimes \!\!\!&\!\! 1 \cr
T^G_{24} \!\!&\!\! = \!\!&\!\! - \!\!&\!\! 1 \!\!\!&\!\! \otimes \!\!\!&\!\! 1 \!\!\!&\!\! \otimes \!\!\!&\!\! 1 \!\!\!&\!\! \otimes \!\!\!&\!\! \si_3 \!\!\!&\!\! \otimes \!\!\!&\!\! \si_2 \!\!\!&\!\! \otimes \!\!\!&\!\! \si_2 \cr
T^G_{34} \!\!&\!\! = \!\!&\!\!   \!\!&\!\! 1 \!\!\!&\!\! \otimes \!\!\!&\!\! 1 \!\!\!&\!\! \otimes \!\!\!&\!\! 1 \!\!\!&\!\! \otimes \!\!\!&\!\! \si_3 \!\!\!&\!\! \otimes \!\!\!&\!\! \si_3 \!\!\!&\!\! \otimes \!\!\!&\!\! 1 \cr
\vp{A^{\Big(}} T^g_1 \!\!&\!\! = \!\!&\!\! -\fr{i}{4} \!\!&\!\! \si_1 \!\!\!&\!\! \otimes \!\!\!&\!\! \si_1 \!\!\!&\!\! \otimes \!\!\!&\!\! 1 \!\!\!&\!\! \otimes \!\!\!&\!\! 1 \!\!\!&\!\! \otimes \!\!\!&\!\! 1 \!\!\!&\!\! \otimes \!\!\!&\!\! \si_2 \cr
      \!\!&\!\!    \!\!&\!\! -\fr{i}{4} \!\!&\!\! \si_2 \!\!\!&\!\! \otimes \!\!\!&\!\! \si_2 \!\!\!&\!\! \otimes \!\!\!&\!\! 1 \!\!\!&\!\! \otimes \!\!\!&\!\! 1 \!\!\!&\!\! \otimes \!\!\!&\!\! 1 \!\!\!&\!\! \otimes \!\!\!&\!\! \si_2 \cr
T^g_2 \!\!&\!\! = \!\!&\!\! -\fr{i}{4} \!\!&\!\! \si_1 \!\!\!&\!\! \otimes \!\!\!&\!\! \si_2 \!\!\!&\!\! \otimes \!\!\!&\!\! 1 \!\!\!&\!\! \otimes \!\!\!&\!\! 1 \!\!\!&\!\! \otimes \!\!\!&\!\! 1 \!\!\!&\!\! \otimes \!\!\!&\!\! 1 \cr
      \!\!&\!\!    \!\!&\!\! +\fr{i}{4} \!\!&\!\! \si_2 \!\!\!&\!\! \otimes \!\!\!&\!\! \si_1 \!\!\!&\!\! \otimes \!\!\!&\!\! 1 \!\!\!&\!\! \otimes \!\!\!&\!\! 1 \!\!\!&\!\! \otimes \!\!\!&\!\! 1 \!\!\!&\!\! \otimes \!\!\!&\!\! 1 \cr
T^g_3 \!\!&\!\! = \!\!&\!\! -\fr{i}{4} \!\!&\!\! \si_3 \!\!\!&\!\! \otimes \!\!\!&\!\! 1 \!\!\!&\!\! \otimes \!\!\!&\!\! 1 \!\!\!&\!\! \otimes \!\!\!&\!\! 1 \!\!\!&\!\! \otimes \!\!\!&\!\! 1 \!\!\!&\!\! \otimes \!\!\!&\!\! \si_2 \cr
      \!\!&\!\!    \!\!&\!\! +\fr{i}{4} \!\!&\!\! 1 \!\!\!&\!\! \otimes \!\!\!&\!\! \si_3 \!\!\!&\!\! \otimes \!\!\!&\!\! 1 \!\!\!&\!\! \otimes \!\!\!&\!\! 1 \!\!\!&\!\! \otimes \!\!\!&\!\! 1 \!\!\!&\!\! \otimes \!\!\!&\!\! \si_2 \cr
T^g_4 \!\!&\!\! = \!\!&\!\! \fr{i}{4} \!\!&\!\! \si_1 \!\!\!&\!\! \otimes \!\!\!&\!\! \si_3 \!\!\!&\!\! \otimes \!\!\!&\!\! 1 \!\!\!&\!\! \otimes \!\!\!&\!\! 1 \!\!\!&\!\! \otimes \!\!\!&\!\! 1 \!\!\!&\!\! \otimes \!\!\!&\!\! \si_2 \cr
           \!\!&\!\!    \!\!&\!\! -\fr{i}{4} \!\!&\!\! \si_1 \!\!\!&\!\! \otimes \!\!\!&\!\! 1 \!\!\!&\!\! \otimes \!\!\!&\!\! 1 \!\!\!&\!\! \otimes \!\!\!&\!\! 1 \!\!\!&\!\! \otimes \!\!\!&\!\! 1 \!\!\!&\!\! \otimes \!\!\!&\!\! \si_2 \cr
T^g_5 \!\!&\!\! = \!\!&\!\! \fr{i}{4} \!\!&\!\! \si_2 \!\!\!&\!\! \otimes \!\!\!&\!\! 1 \!\!\!&\!\! \otimes \!\!\!&\!\! 1 \!\!\!&\!\! \otimes \!\!\!&\!\! 1 \!\!\!&\!\! \otimes \!\!\!&\!\! 1 \!\!\!&\!\! \otimes \!\!\!&\!\! 1 \cr
           \!\!&\!\!    \!\!&\!\! -\fr{i}{4} \!\!&\!\! \si_2 \!\!\!&\!\! \otimes \!\!\!&\!\! \si_3 \!\!\!&\!\! \otimes \!\!\!&\!\! 1 \!\!\!&\!\! \otimes \!\!\!&\!\! 1 \!\!\!&\!\! \otimes \!\!\!&\!\! 1 \!\!\!&\!\! \otimes \!\!\!&\!\! 1 \cr
T^g_6 \!\!&\!\! = \!\!&\!\! -\fr{i}{4} \!\!&\!\! 1 \!\!\!&\!\! \otimes \!\!\!&\!\! \si_1 \!\!\!&\!\! \otimes \!\!\!&\!\! 1 \!\!\!&\!\! \otimes \!\!\!&\!\! 1 \!\!\!&\!\! \otimes \!\!\!&\!\! 1 \!\!\!&\!\! \otimes \!\!\!&\!\! \si_2 \cr
           \!\!&\!\!    \!\!&\!\! +\fr{i}{4} \!\!&\!\! \si_3 \!\!\!&\!\! \otimes \!\!\!&\!\! \si_1 \!\!\!&\!\! \otimes \!\!\!&\!\! 1 \!\!\!&\!\! \otimes \!\!\!&\!\! 1 \!\!\!&\!\! \otimes \!\!\!&\!\! 1 \!\!\!&\!\! \otimes \!\!\!&\!\! \si_2 \cr
T^g_7 \!\!&\!\! = \!\!&\!\! -\fr{i}{4} \!\!&\!\! \si_3 \!\!\!&\!\! \otimes \!\!\!&\!\! \si_2 \!\!\!&\!\! \otimes \!\!\!&\!\! 1 \!\!\!&\!\! \otimes \!\!\!&\!\! 1 \!\!\!&\!\! \otimes \!\!\!&\!\! 1 \!\!\!&\!\! \otimes \!\!\!&\!\! 1 \cr
           \!\!&\!\!    \!\!&\!\! +\fr{i}{4} \!\!&\!\! 1 \!\!\!&\!\! \otimes \!\!\!&\!\! \si_2 \!\!\!&\!\! \otimes \!\!\!&\!\! 1 \!\!\!&\!\! \otimes \!\!\!&\!\! 1 \!\!\!&\!\! \otimes \!\!\!&\!\! 1 \!\!\!&\!\! \otimes \!\!\!&\!\! 1 \cr
T^g_8 \!\!&\!\! = \!\!&\!\! -\fr{i}{4 \sqrt{3}} \!\!&\!\! \si_3 \!\!\!&\!\! \otimes \!\!\!&\!\! 1 \!\!\!&\!\! \otimes \!\!\!&\!\! 1 \!\!\!&\!\! \otimes \!\!\!&\!\! 1 \!\!\!&\!\! \otimes \!\!\!&\!\! 1 \!\!\!&\!\! \otimes \!\!\!&\!\! \si_2 \cr
           \!\!&\!\!    \!\!&\!\! -\fr{i}{4 \sqrt{3}} \!\!&\!\! 1 \!\!\!&\!\! \otimes \!\!\!&\!\! \si_3 \!\!\!&\!\! \otimes \!\!\!&\!\! 1 \!\!\!&\!\! \otimes \!\!\!&\!\! 1 \!\!\!&\!\! \otimes \!\!\!&\!\! 1 \!\!\!&\!\! \otimes \!\!\!&\!\! \si_2 \cr
           \!\!&\!\!    \!\!&\!\! +\fr{i}{2 \sqrt{3}} \!\!&\!\! \si_3 \!\!\!&\!\! \otimes \!\!\!&\!\! \si_3 \!\!\!&\!\! \otimes \!\!\!&\!\! 1 \!\!\!&\!\! \otimes \!\!\!&\!\! 1 \!\!\!&\!\! \otimes \!\!\!&\!\! 1 \!\!\!&\!\! \otimes \!\!\!&\!\! \si_2 \cr
\vp{A^{\Big(}} T^W_1 \!\!&\!\! = \!\!&\!\! -\fr{i}{4} \!\!&\!\! 1 \!\!\!&\!\! \otimes \!\!\!&\!\! 1 \!\!\!&\!\! \otimes \!\!\!&\!\! \si_1 \!\!\!&\!\! \otimes \!\!\!&\!\! 1 \!\!\!&\!\! \otimes \!\!\!&\!\! 1 \!\!\!&\!\! \otimes \!\!\!&\!\! \si_2 \cr
            \!\!&\!\!    \!\!&\!\! -\fr{i}{4} \!\!&\!\! 1 \!\!\!&\!\! \otimes \!\!\!&\!\! 1 \!\!\!&\!\! \otimes \!\!\!&\!\! \si_1 \!\!\!&\!\! \otimes \!\!\!&\!\! \si_3 \!\!\!&\!\! \otimes \!\!\!&\!\! 1 \!\!\!&\!\! \otimes \!\!\!&\!\! \si_2 \cr
T^W_2 \!\!&\!\! = \!\!&\!\! \fr{i}{4} \!\!&\!\! 1 \!\!\!&\!\! \otimes \!\!\!&\!\! 1 \!\!\!&\!\! \otimes \!\!\!&\!\! \si_2 \!\!\!&\!\! \otimes \!\!\!&\!\! 1 \!\!\!&\!\! \otimes \!\!\!&\!\! 1 \!\!\!&\!\! \otimes \!\!\!&\!\! 1 \cr
            \!\!&\!\!    \!\!&\!\! \fr{i}{4} \!\!&\!\! 1 \!\!\!&\!\! \otimes \!\!\!&\!\! 1 \!\!\!&\!\! \otimes \!\!\!&\!\! \si_2 \!\!\!&\!\! \otimes \!\!\!&\!\! \si_3 \!\!\!&\!\! \otimes \!\!\!&\!\! 1 \!\!\!&\!\! \otimes \!\!\!&\!\! 1 \cr
T^W_3 \!\!&\!\! = \!\!&\!\! -\fr{i}{4} \!\!&\!\! 1 \!\!\!&\!\! \otimes \!\!\!&\!\! 1 \!\!\!&\!\! \otimes \!\!\!&\!\! \si_3 \!\!\!&\!\! \otimes \!\!\!&\!\! \si_3 \!\!\!&\!\! \otimes \!\!\!&\!\! 1 \!\!\!&\!\! \otimes \!\!\!&\!\! \si_2 \cr
            \!\!&\!\!    \!\!&\!\! -\fr{i}{4} \!\!&\!\! 1 \!\!\!&\!\! \otimes \!\!\!&\!\! 1 \!\!\!&\!\! \otimes \!\!\!&\!\! \si_3 \!\!\!&\!\! \otimes \!\!\!&\!\! 1 \!\!\!&\!\! \otimes \!\!\!&\!\! 1 \!\!\!&\!\! \otimes \!\!\!&\!\! \si_2 \cr
\vp{A^{\Big(}}  T^Y \!\!&\!\! = \!\!&\!\! \fr{i}{4} \!\!&\!\! 1 \!\!\!&\!\! \otimes \!\!\!&\!\! 1 \!\!\!&\!\! \otimes \!\!\!&\!\! \si_3 \!\!\!&\!\! \otimes \!\!\!&\!\! \si_3 \!\!\!&\!\! \otimes \!\!\!&\!\! 1 \!\!\!&\!\! \otimes \!\!\!&\!\! \si_2 \cr
           \!\!&\!\!    \!\!&\!\! -\fr{i}{4} \!\!&\!\! 1 \!\!\!&\!\! \otimes \!\!\!&\!\! 1 \!\!\!&\!\! \otimes \!\!\!&\!\! \si_3 \!\!\!&\!\! \otimes \!\!\!&\!\! 1 \!\!\!&\!\! \otimes \!\!\!&\!\! 1 \!\!\!&\!\! \otimes \!\!\!&\!\! \si_2 \cr
           \!\!&\!\!    \!\!&\!\! +\fr{i}{6} \!\!&\!\! \si_3 \!\!\!&\!\! \otimes \!\!\!&\!\! 1 \!\!\!&\!\! \otimes \!\!\!&\!\! 1 \!\!\!&\!\! \otimes \!\!\!&\!\! 1 \!\!\!&\!\! \otimes \!\!\!&\!\! 1 \!\!\!&\!\! \otimes \!\!\!&\!\! \si_2 \cr
           \!\!&\!\!    \!\!&\!\! +\fr{i}{6} \!\!&\!\! 1 \!\!\!&\!\! \otimes \!\!\!&\!\! \si_3 \!\!\!&\!\! \otimes \!\!\!&\!\! 1 \!\!\!&\!\! \otimes \!\!\!&\!\! 1 \!\!\!&\!\! \otimes \!\!\!&\!\! 1 \!\!\!&\!\! \otimes \!\!\!&\!\! \si_2 \cr
           \!\!&\!\!    \!\!&\!\! +\fr{i}{6} \!\!&\!\! \si_3 \!\!\!&\!\! \otimes \!\!\!&\!\! \si_3 \!\!\!&\!\! \otimes \!\!\!&\!\! 1 \!\!\!&\!\! \otimes \!\!\!&\!\! 1 \!\!\!&\!\! \otimes \!\!\!&\!\! 1 \!\!\!&\!\! \otimes \!\!\!&\!\! \si_2 \cr
\end{array}
\;\;\;\;\;\;\;\;\;\;\;\;\;
\ps=
\lb \!\!
\begin{array}{c}
\nu \cr e \cr u^r \cr d^r \cr u^g \cr d^g \cr u^b \cr d^b \cr 
\end{array}
\!\! \rb
\begin{array}{c} 
\left\{
\;
\lb \!\!
\begin{array}{c}
e_{Lr}^{\wedge} \cr e_{Li}^{\wedge} \cr e_{Lr}^{\vee} \cr e_{Li}^{\vee} \cr
e_{Rr}^{\wedge} \cr e_{Ri}^{\wedge} \cr e_{Rr}^{\vee} \cr e_{Ri}^{\vee} \cr
\end{array}
\!\! \rb
\right.
\cr
\vp{a} \cr \vp{a} \cr
\vp{a} \cr \vp{a} \cr
\vp{a} \cr
\end{array}
\label{sm}
\eeq

These eighteen $\mathbb{R}(64)$ matrices, representing gravitational and Standard Model generators acting on each other via their Lie bracket and on a generation of fermion components in a $64^\mathbb{R}$ column, constitute the conventional description of the gravitational and Standard Model algebra. Any successful unified description of gravity and the Standard Model must describe how these algebraic elements embed in a larger algebra.

\section{Unification in a GraviGUT}

The gravitational and Standard Model Lie algebra, $su(3) \!\oplus\! su(2)_L \!\oplus\! u(1)_Y \!\oplus\! spin(1,3)$, a subalgebra of $spin(11,3)$, acts on a generation of Standard Model fermions in a specific representation, (\ref{sm}), corresponding to the particle interactions of bosons and fermions. Remarkably, this $64^\mathbb{R}$ representation space of fermions is equal to a Majorana-Weyl spinor of $spin(11,3)$. To establish this equivalence, we build a matrix representation of $spin(11,3)$ from a representation of the $Cl(11,3)$ Clifford algebra. Specifically, we choose this {\it very nice} representation of fourteen $Cl^1(11,3)$ Clifford basis vectors as real $\mathbb{R}(128)$ matrices:
\beq
\renewcommand{\arraystretch}{1.1}
\begin{array}{rcrccccccccccccc}
\Ga_1 \!\!&\!\! = \!\!&\!\! i \!\!&\!\! \si_1 \!\!\!&\!\! \otimes \!\!\!&\!\! 1 \!\!\!&\!\! \otimes \!\!\!&\!\! 1 \!\!\!&\!\! \otimes \!\!\!&\!\! \si_2 \!\!\!&\!\! \otimes \!\!\!&\!\! \si_2 \!\!\!&\!\! \otimes \!\!\!&\!\! \si_1 \!\!\!&\!\! \otimes \!\!\!&\!\! \si_2 \cr
\Ga_2 \!\!&\!\! = \!\!&\!\! -i \!\!&\!\! \si_1 \!\!\!&\!\! \otimes \!\!\!&\!\! 1 \!\!\!&\!\! \otimes \!\!\!&\!\! 1 \!\!\!&\!\! \otimes \!\!\!&\!\! \si_2 \!\!\!&\!\! \otimes \!\!\!&\!\! \si_2 \!\!\!&\!\! \otimes \!\!\!&\!\! \si_2 \!\!\!&\!\! \otimes \!\!\!&\!\! 1 \cr
\Ga_3 \!\!&\!\! = \!\!&\!\! i \!\!&\!\! \si_1 \!\!\!&\!\! \otimes \!\!\!&\!\! 1 \!\!\!&\!\! \otimes \!\!\!&\!\! 1 \!\!\!&\!\! \otimes \!\!\!&\!\! \si_2 \!\!\!&\!\! \otimes \!\!\!&\!\! \si_2 \!\!\!&\!\! \otimes \!\!\!&\!\! \si_3 \!\!\!&\!\! \otimes \!\!\!&\!\! \si_2 \cr
\Ga_4 \!\!&\!\! = \!\!&\!\!  \!\!&\!\! \si_1 \!\!\!&\!\! \otimes \!\!\!&\!\! 1 \!\!\!&\!\! \otimes \!\!\!&\!\! 1 \!\!\!&\!\! \otimes \!\!\!&\!\! \si_2 \!\!\!&\!\! \otimes \!\!\!&\!\! \si_1 \!\!\!&\!\! \otimes \!\!\!&\!\! 1 \!\!\!&\!\! \otimes \!\!\!&\!\! \si_2 \cr
\Ga_5 \!\!&\!\! = \!\!&\!\!  \!\!&\!\! \si_1 \!\!\!&\!\! \otimes \!\!\!&\!\! 1 \!\!\!&\!\! \otimes \!\!\!&\!\! 1 \!\!\!&\!\! \otimes \!\!\!&\!\! \si_1 \!\!\!&\!\! \otimes \!\!\!&\!\! 1 \!\!\!&\!\! \otimes \!\!\!&\!\! 1 \!\!\!&\!\! \otimes \!\!\!&\!\! 1 \cr
\Ga_6 \!\!&\!\! = \!\!&\!\!  \!\!&\!\! \si_1 \!\!\!&\!\! \otimes \!\!\!&\!\! 1 \!\!\!&\!\! \otimes \!\!\!&\!\! 1 \!\!\!&\!\! \otimes \!\!\!&\!\! \si_2 \!\!\!&\!\! \otimes \!\!\!&\!\! \si_3 \!\!\!&\!\! \otimes \!\!\!&\!\! 1 \!\!\!&\!\! \otimes \!\!\!&\!\! \si_2 \cr
\Ga_7 \!\!&\!\! = \!\!&\!\!  \!\!&\!\! \si_1 \!\!\!&\!\! \otimes \!\!\!&\!\! 1 \!\!\!&\!\! \otimes \!\!\!&\!\! 1 \!\!\!&\!\! \otimes \!\!\!&\!\! \si_3 \!\!\!&\!\! \otimes \!\!\!&\!\! 1 \!\!\!&\!\! \otimes \!\!\!&\!\! 1 \!\!\!&\!\! \otimes \!\!\!&\!\! 1 \cr
\Ga_8 \!\!&\!\! = \!\!&\!\! - \!\!&\!\! \si_2 \!\!\!&\!\! \otimes \!\!\!&\!\! 1 \!\!\!&\!\! \otimes \!\!\!&\!\! 1 \!\!\!&\!\! \otimes \!\!\!&\!\! 1 \!\!\!&\!\! \otimes \!\!\!&\!\! 1 \!\!\!&\!\! \otimes \!\!\!&\!\! 1 \!\!\!&\!\! \otimes \!\!\!&\!\! \si_2 \cr
\Ga_9 \!\!&\!\! = \!\!&\!\!  \!\!&\!\! \si_2 \!\!\!&\!\! \otimes \!\!\!&\!\! \si_3 \!\!\!&\!\! \otimes \!\!\!&\!\! \si_2 \!\!\!&\!\! \otimes \!\!\!&\!\! 1 \!\!\!&\!\! \otimes \!\!\!&\!\! \si_2 \!\!\!&\!\! \otimes \!\!\!&\!\! \si_2 \!\!\!&\!\! \otimes \!\!\!&\!\! \si_1 \cr
\Ga_{10} \!\!&\!\! = \!\!&\!\! - \!\!&\!\! \si_2 \!\!\!&\!\! \otimes \!\!\!&\!\! 1 \!\!\!&\!\! \otimes \!\!\!&\!\! \si_2 \!\!\!&\!\! \otimes \!\!\!&\!\! 1 \!\!\!&\!\! \otimes \!\!\!&\!\! \si_2 \!\!\!&\!\! \otimes \!\!\!&\!\! \si_2 \!\!\!&\!\! \otimes \!\!\!&\!\! \si_3 \cr
\Ga_{11} \!\!&\!\! = \!\!&\!\! - \!\!&\!\! \si_2 \!\!\!&\!\! \otimes \!\!\!&\!\! \si_2 \!\!\!&\!\! \otimes \!\!\!&\!\! 1 \!\!\!&\!\! \otimes \!\!\!&\!\! 1 \!\!\!&\!\! \otimes \!\!\!&\!\! \si_2 \!\!\!&\!\! \otimes \!\!\!&\!\! \si_2 \!\!\!&\!\! \otimes \!\!\!&\!\! \si_1 \cr
\Ga_{12} \!\!&\!\! = \!\!&\!\!  \!\!&\!\! \si_2 \!\!\!&\!\! \otimes \!\!\!&\!\! \si_2 \!\!\!&\!\! \otimes \!\!\!&\!\! \si_3 \!\!\!&\!\! \otimes \!\!\!&\!\! 1 \!\!\!&\!\! \otimes \!\!\!&\!\! \si_2 \!\!\!&\!\! \otimes \!\!\!&\!\! \si_2 \!\!\!&\!\! \otimes \!\!\!&\!\! \si_3 \cr
\Ga_{13} \!\!&\!\! = \!\!&\!\!  \!\!&\!\! \si_2 \!\!\!&\!\! \otimes \!\!\!&\!\! \si_1 \!\!\!&\!\! \otimes \!\!\!&\!\! \si_2 \!\!\!&\!\! \otimes \!\!\!&\!\! 1 \!\!\!&\!\! \otimes \!\!\!&\!\! \si_2 \!\!\!&\!\! \otimes \!\!\!&\!\! \si_2 \!\!\!&\!\! \otimes \!\!\!&\!\! \si_1 \cr
\Ga_{14} \!\!&\!\! = \!\!&\!\! - \!\!&\!\! \si_2 \!\!\!&\!\! \otimes \!\!\!&\!\! \si_2 \!\!\!&\!\! \otimes \!\!\!&\!\! \si_1 \!\!\!&\!\! \otimes \!\!\!&\!\! 1 \!\!\!&\!\! \otimes \!\!\!&\!\! \si_2 \!\!\!&\!\! \otimes \!\!\!&\!\! \si_2 \!\!\!&\!\! \otimes \!\!\!&\!\! \si_3 \cr
\end{array}
\label{Ga}
\eeq
These satisfy the Clifford algebra basis vector identity
$$
\Ga_i \Ga_j = \ha (\Ga_i \Ga_j + \Ga_j \Ga_i) + \ha (\Ga_i \Ga_j - \Ga_j \Ga_i) = \Ga_i \cdot \Ga_j \,+\, \Ga_i \times \Ga_j = \et_{ij} 1 + \Ga_{ij}
$$
in which the Clifford product of these generators is equivalent to the product of the representative matrices. The $\et_{ij}$ is the Minkowski metric of signature $(11,3)$. The $\Ga_{ij} = \Ga_i \Ga_j$, with $i<j$, are nintey-one independent bivector basis generators, spanning $Cl^2(11,3)$. As $\mathbb{R}(128)$ matrices, these bivectors are block diagonal,
\beq
\Ga_{ij} =
\lb
\begin{array}{cc}
\Ga^+_{ij} & 0 \cr
0 & \Ga^-_{ij} \cr
\end{array}
\rb
\label{bGa}
\eeq
with $\Ga^+_{ij}$ and $\Ga^-_{ij}$ the $\mathbb{R}(64)$ basis matrix generators spanning the positive and negative chiral representations of $spin(11,3)$. The $spin(11,3)$ Lie algebra generators may be represented by either chiral set of matricies, $\Ga^{\pm}_{ij}$, or by the larger $\Ga_{ij}$. The $\Ga^+_{ij}$ matrices act on a real $64^\mathbb{R}_{S+}$ column matrix -- a positive chiral Majorana-Weyl $spin(11,3)$ spinor.

The gravitational and Standard Model Lie algebra is a subalgebra of $spin(11,3)$, spanned by eighteen basis generators,
\beq
\renewcommand{\arraystretch}{1.25}
\begin{array}{rcl}
T^G_{23} \!\!&\!\! = \!\!&\!\! \Ga^+_{2 \, 3} \cr
T^G_{13} \!\!&\!\! = \!\!&\!\! \Ga^+_{1 \, 3} \cr
T^G_{12} \!\!&\!\! = \!\!&\!\! \Ga^+_{1 \, 2} \cr
T^G_{14} \!\!&\!\! = \!\!&\!\! \Ga^+_{1 \, 4} \cr
T^G_{24} \!\!&\!\! = \!\!&\!\! \Ga^+_{2 \, 4} \cr
T^G_{34} \!\!&\!\! = \!\!&\!\! \Ga^+_{3 \, 4} \cr
\p{a} \!\!&\!\! \p{a} \!\!&\!\! \p{a} \cr
\p{a} \!\!&\!\! \p{a} \!\!&\!\! \p{a} \cr
\end{array}
\;\;\;
\begin{array}{rcl}
T^g_1 \!\!&\!\! = \!\!&\!\! \fr{1}{4}  \Ga^+_{9 \, 12} - \fr{1}{4}  \Ga^+_{10 \, 11}  \cr
T^g_2 \!\!&\!\! = \!\!&\!\! -\fr{1}{4}  \Ga^+_{9 \, 11} - \fr{1}{4}  \Ga^+_{10 \, 12}  \cr
T^g_3 \!\!&\!\! = \!\!&\!\! -\fr{1}{4}  \Ga^+_{9 \, 10} + \fr{1}{4}  \Ga^+_{11 \, 12}  \cr
T^g_4 \!\!&\!\! = \!\!&\!\! -\fr{1}{4}  \Ga^+_{9 \, 14} + \fr{1}{4}  \Ga^+_{10 \, 13}  \cr
T^g_5 \!\!&\!\! = \!\!&\!\! \fr{1}{4}  \Ga^+_{9 \, 13} + \fr{1}{4}  \Ga^+_{10 \, 14}  \cr
T^g_6 \!\!&\!\! = \!\!&\!\! \fr{1}{4}  \Ga^+_{11 \, 14} - \fr{1}{4}  \Ga^+_{12 \, 13}  \cr
T^g_7 \!\!&\!\! = \!\!&\!\! -\fr{1}{4}  \Ga^+_{11 \, 13} - \fr{1}{4}  \Ga^+_{12 \, 14}  \cr
T^g_8 \!\!&\!\! = \!\!&\!\! -\fr{1}{4 \sqrt{3}}  \Ga^+_{9 \, 10} - \fr{1}{4 \sqrt{3}}  \Ga^+_{11 \, 12} + \fr{1}{2 \sqrt{3}}  \Ga^+_{13 \, 14} \cr
\end{array}
\!\!\!\!\!\!\!\!\!\!\!\!\!\!\!\!\!\!\!\!\!\!\!\!\!\!\!\!\!\!\!\!\!\!\!\!\!\!
\begin{array}{rcl}
T^W_1 \!\!&\!\! = \!\!&\!\! \fr{1}{4}  \Ga^+_{5 \, 8} - \fr{1}{4}  \Ga^+_{6 \, 7}  \cr
T^W_2 \!\!&\!\! = \!\!&\!\! -\fr{1}{4}  \Ga^+_{5 \, 7} - \fr{1}{4}  \Ga^+_{6 \, 8}  \cr
T^W_3 \!\!&\!\! = \!\!&\!\! -\fr{1}{4}  \Ga^+_{5 \, 6} + \fr{1}{4}  \Ga^+_{7 \, 8}  \cr
\p{a} \!\!&\!\! \p{a} \!\!&\!\! \p{a} \cr
T^Y \!\!&\!\! = \!\!&\!\! \fr{1}{4}  \Ga^+_{5 \, 6} + \fr{1}{4}  \Ga^+_{7 \, 8} + \fr{1}{6}  \Ga^+_{9 \, 10} \cr
 \!\!&\!\!  \!\!&\!\!  + \fr{1}{6}  \Ga^+_{11 \, 12} + \fr{1}{6}  \Ga^+_{13 \, 14} \cr
\p{a} \!\!&\!\! \p{a} \!\!&\!\! \p{a} \cr
\p{a} \!\!&\!\! \p{a} \!\!&\!\! \p{a} \cr
\end{array}
\label{ssm}
\eeq
Because of our {\it very nice} Clifford algebra representation, (\ref{Ga}), these eighteen $\mathbb{R}(64)$ matrix generators, (\ref{ssm}), are {\it numerically identical} to the conventional gravitational and Standard Model basis generator matrices, (\ref{sm}). With this identification, we also see directly that a generation of fermions, in a $64^\mathbb{R}$ of (\ref{sm}), is identified as a positive chiral Majorana-Weyl $spin(11,3)$ spinor, $64^\mathbb{R}_{S+}$, with identical components. This embedding of the gravitational and Standard Model algebra in a $spin(11,3)$ GraviGUT is in agreement with the recent work of Nesti and Percacci,\cite{Percacci} and with a unified GraviGUT action.\cite{LSS} It is worth noting that the embedding of the gravitational and Standard Model Lie algebra in $spin(11,3)$ expressed in (\ref{ssm}) in terms of bivector generators is valid regardless of the choice of matrix representation. Nevertheless, the main value of the present work is in describing this embedding directly and explicitly, as we have here, via a convenient choice of matrix representation. The Clifford basis vector matrices, (\ref{Ga}), are essentially the square root of the conventional Standard Model generators.

This Clifford basis vector representation, (\ref{Ga}), is useful not only for matching the Standard Model generators, but also for matching the $su(4) \!\oplus\! su(2)_L \!\oplus\! su(2)_R$ basis generators of the Pati-Salam GUT. For these, the seven standard $su(4)$ basis generators not in the strong $su(3)$ subalgebra, and the three $su(2)_R$ basis generators, are:
\beq
\renewcommand{\arraystretch}{1.25}
\begin{array}{rcl}
T^g_9 \!\!&\!\! = \!\!&\!\! \fr{1}{4}  \Ga^+_{11 \, 14} + \fr{1}{4}  \Ga^+_{12 \, 13}  \cr
T^g_{10} \!\!&\!\! = \!\!&\!\! \fr{1}{4}  \Ga^+_{11 \, 13} - \fr{1}{4}  \Ga^+_{12 \, 14}  \cr
T^g_{11} \!\!&\!\! = \!\!&\!\! \fr{1}{4}  \Ga^+_{9 \, 14} + \fr{1}{4}  \Ga^+_{10 \, 13}  \cr
T^g_{12} \!\!&\!\! = \!\!&\!\! \fr{1}{4}  \Ga^+_{9 \, 13} - \fr{1}{4}  \Ga^+_{10 \, 14}  \cr
T^g_{13} \!\!&\!\! = \!\!&\!\! \fr{1}{4}  \Ga^+_{9 \, 12} + \fr{1}{4}  \Ga^+_{10 \, 11}  \cr
T^g_{14} \!\!&\!\! = \!\!&\!\! \fr{1}{4}  \Ga^+_{9 \, 11} - \fr{1}{4}  \Ga^+_{10 \, 12}  \cr
T^g_{15} \!\!&\!\! = \!\!&\!\! \fr{1}{2 \sqrt{6}}  \Ga^+_{9 \, 10} + \fr{1}{2 \sqrt{6}} \Ga^+_{11 \, 12} + \fr{1}{2 \sqrt{6}} \Ga^+_{13 \, 14} \cr
\end{array}
\!\!\!\!\!\!\!\!\!\!\!\!\!\!\!\!\!\!\!\!\!\!
\begin{array}{rcl}
T^{W'}_1 \!\!&\!\! = \!\!&\!\! \fr{1}{4}  \Ga^+_{5 \, 8} + \fr{1}{4}  \Ga^+_{6 \, 7}  \cr
T^{W'}_2 \!\!&\!\! = \!\!&\!\! -\fr{1}{4}  \Ga^+_{5 \, 7} + \fr{1}{4}  \Ga^+_{6 \, 8}  \cr
T^{W'}_3 \!\!&\!\! = \!\!&\!\! \fr{1}{4}  \Ga^+_{5 \, 6} + \fr{1}{4}  \Ga^+_{7 \, 8}  \cr
\p{a} \!\!&\!\! \p{a} \!\!&\!\! \p{a} \cr
\p{a} \!\!&\!\! \p{a} \!\!&\!\! \p{a} \cr
\p{a} \!\!&\!\! \p{a} \!\!&\!\! \p{a} \cr
\p{a} \!\!&\!\! \p{a} \!\!&\!\! \p{a} \cr
\end{array}
\label{ps}
\eeq
In the Pati-Salam GUT, these twenty-one basis generators, spanning an $\mathbb{R}(64)$ matrix representation of $su(4) \!\oplus\! su(2)_L \!\oplus\! su(2)_R$, act on fermions as in (\ref{sm}). The twelve basis generators of the Standard Model in (\ref{ssm}) are a subset of these twenty-one. Thus, the Standard Model Lie algebra embeds in $su(4) \!\oplus\! su(2)_L \!\oplus\! su(2)_R$, which is a subalgebra of $spin(10)$, which is a subalgebra of $spin(11,3)$. The six basis generators of gravitational $spin(1,3)$ in (\ref{ssm}) are a subset of the ninety-one $spin(11,3)$ basis generators. The other forty basis generators in $spin(11,3)$, other than the forty-five of $spin(10)$ and six of $spin(1,3)$, may be used to describe a {\it frame-Higgs} multiplet.

\subsection{Frame-Higgs and the Unified Bosonic Connection}

In modern formulations of gravity, the {\it frame}, locally $\f{e} = \f{dx}^\mu (e_\mu)^a \ga_a$, in which $\ga_a$ are four $Cl^1(1,3)$ Clifford basis vectors, (\ref{cl13}), describes a rest frame at every point of spacetime. The frame relates a set of four orthonormal basis vectors, $\ve{e}{}_a = (e^-_a)^\mu \ve{\pa}{}_\mu$, at each spacetime point to the $\ga_a$ spanning a local Minkowski spacetime, determining the spacetime metric, $g_{\mu \nu} = (e_\mu)^a \eta_{ab} (e_\nu)^b$. The {\it spin connection}, locally $\f{\om} = \ha \f{\om}{}^{ab} \ga_{ab}$, valued in $spin(1,3) = Cl^2(1,3)$, acts on the frame as a 4-vector, and describes how it changes over spacetime. The covariant derivative of the frame is the gravitational torsion 2-form,
\beq
\ff{T} = \f{D} \f{e} = \f{d} \f{e} + \f{\om} \times \f{e} = \f{d} \f{e} + \ha \f{\om} \f{e} + \ha \f{e} \f{\om}
= ( \f{d} \f{e}^a + \f{\om}{}^a{}_b \f{e}{}^b ) \ga_a 
\eeq
which is usually zero, allowing the spin connection to be determined from the frame. In such expressions a wedge is always implied between differential forms, with underlines used to designate form grade, and we use the Clifford product -- equivalent to the matrix product between representative matrices.

In the $spin(10)$ GUT there is usually a scalar Higgs multiplet that is a 10-vector, $\ph = \ph^P \Ga_P$. Using the $Cl(10)$ Clifford product, the covariant derivative of such a Higgs is
$$
\f{D} \ph = \f{d} \ph + 2 \f{A} \times \ph = \f{d} \ph + \f{A} \ph - \ph \f{A}
= ( \f{d} \ph^P + 2 \f{A}{}^P{}_Q \ph^Q ) \Ga_P 
$$
We can combine all bosonic fields -- spin connection, frame, Higgs, and gauge fields -- in the {\it unified bosonic connection} of $spin(11,3)$,
\beq
\f{H} = \ha \f{\om} + \fr{1}{4} \f{e} \ph + \f{A} = \fr{1}{4} \f{\om}{}^{ab} \Ga_{ab} + \fr{1}{4} \f{e}{}^a \ph^P \Ga_{aP} + \ha \f{A}{}^{PQ} \Ga_{PQ}
\label{ubc}
\eeq
in which we have substituted $\Ga_a$'s for $\ga_a$'s in the gravitational frame, $\f{e}$, and spin connection, $\f{\om}$. In this way, the gravitational frame and a Higgs naturally emerge as the {\it frame-Higgs} part, $\f{e} \ph$, of a $spin(11,3)$ connection. The curvature of this unified bosonic connection is
\beq
\ff{F} = \f{d} \f{H} + \f{H} \f{H} = \ha \lp \ff{R} - \fr{1}{8} \f{e} \f{e} \ph^2 \rp + \fr{1}{4} \lp \ff{T} \ph - \f{e} \f{D} \ph \rp + \ff{F}{}_A
\label{curvature}
\eeq
in which $\ff{R} = \f{d} \f{\om} + \ha \f{\om} \f{\om}$ is the Riemann curvature 2-form, we recognize $\ff{T}$ and $\f{D} \ph$ as the torsion and covariant derivative of the Higgs, and $\ff{F}{}_A = \f{d} \f{A} + \f{A} \f{A}$ is the curvature of the $spin(10)$ gauge field.

The dynamical symmetry breaking of $spin(11,3)$ to gravitational $spin(1,3)$ and $spin(10)$ is described in more detail in \cite{LSS}. To summarize, the frame-Higgs, $\f{e}\ph$, becomes nonzero over all of spacetime, reducing the structure group to that of gravity and the Standard Model, with the frame, $\f{e}$, describing a de Sitter cosmology, and the Higgs, $\ph \simeq \ph_0$, having a nonzero vacuum expectation value. Gauge fields interact with $\ph_0$, becoming massive, and this Higgs background also gives masses to fermions via the covariant Dirac derivative,
\beq
\f{D} \ps = \f{d} \ps + \f{H} \ps = (\f{d} + \ha \f{\om} + \fr{1}{4} \f{e} \ph + \f{A} ) \ps
\label{dirac}
\eeq
Interactions between elementary particles correspond to the geometric and algebraic structure expressed in the curvature (\ref{curvature}) and covariant Dirac derivative (\ref{dirac}) as they appear in an action.

\subsection{GraviGUT Algebra}

The structure of the $spin(11,3)$ GraviGUT Lie algebra, corresponding to interactions between bosons, is described explicitly by the Lie bracket between generators,
\beq
[\Ga_{ij}, \Ga_{kl}] = 2 \eta_{jk} \Ga_{il} - 2 \eta_{jl} \Ga_{ik} + 2 \eta_{il} \Ga_{jk} - 2 \eta_{ik} \Ga_{jl}
\label{s113}
\eeq
Elements of $spin(11,3)$ also act as linear operators on the Majorana-Weyl spinor representation space, shuffling the components of fermions in a $64^\mathbb{R}_{S+}$, corresponding to interactions between bosons and fermions. If we write such a spinor as $\ps = \ps^\io Q_\io$, in terms of coefficients, $\ps^\io$, multiplying canonical basis column matrix {\it spinor generators}, $Q_\io$, then the multiplicative operation of $spin(11,3)$ generators on these can be used to define a generalized bracket,
\beq
[\Ga_{ij}, Q_\io] = \Ga^+_{ij} Q_\io = Q_\ka (\Ga^+_{ij})^\ka_{\p{\ka} \io}
\label{s113s}
\eeq
with $(\Ga^+_{ij})^\ka_{\p{\ka} \io}$ the numerical components of the $\Ga^+_{ij}$ matrices. Together, the $spin(11,3)$ Lie algebra, (\ref{s113}), and its action on Majorana-Weyl spinors, (\ref{s113s}), constitute the {\it GraviGUT algebra}. The word ``algebra'' is used here in a generalized sense, even though the bracket between two spinors may be undefined. If we define such a bracket trivially,
\beq
[Q_\io, Q_\ka] = 0
\label{s113t}
\eeq
and presume $[Q_\io, \Ga_{ij}] = - [\Ga_{ij}, Q_\io]$, then these brackets, (\ref{s113}-\ref{s113t}), are the Lie brackets of a Lie algebra -- completing the GraviGUT algebra -- in which spinors are an ideal. Similarly, it is possible to construct the Standard Model algebra by using the Standard Model Lie algebra generators in (\ref{s113}) and the matrix representation, (\ref{sm}), in (\ref{s113s}), with the $64^\mathbb{R}$ as an ideal. The Standard Model algebra is thus a subalgebra of the GraviGUT algebra.

\section{Unification in $E_8$}

Remarkably, the GraviGUT algebra, defined by (\ref{s113}) and (\ref{s113s}), is a subalgebra of the quaternionic real form of the $E_8$ Lie algebra, $E_{8(-24)}$. This can be seen immediately by using an explicit construction of $E_{8(-24)}$ involving a $spin(12,4)$ subalgebra acting on a $128^\mathbb{R}_{S+}$.

To construct a convenient positive chiral real matrix representation of \linebreak $spin(12,4)$ we first choose a set of representative $\mathbb{R}(256)$ matrices for the sixteen $Cl(12,4)$ Clifford basis vectors:
\beq
\renewcommand{\arraystretch}{1.1}
\begin{array}{rcrccc}
\Ga'_i \!\!&\!\! = \!\!&\!\!  \!\!&\!\! \si_1 \!\!\!&\!\! \otimes \!\!\!&\!\! \Ga_i  \cr
\Ga'_{15} \!\!&\!\! = \!\!&\!\!  \!\!&\!\! \si_1 \!\!\!&\!\! \otimes \!\!\!&\!\! \Ga  \cr
\Ga'_{16} \!\!&\!\! = \!\!&\!\! -i \!\!&\!\! \si_2 \!\!\!&\!\! \otimes \!\!\!&\!\! 1  \cr
\end{array}
\eeq
in which the $\Ga_i$ are the fourteen $\mathbb{R}(128)$ matrices in (\ref{Ga}), and
$$
\Ga = \Ga_{1} \Ga_{2} \Ga_{3} \Ga_{4} \Ga_{5} \Ga_{6} \Ga_{7} \Ga_{8} \Ga_{9} \Ga_{10} \Ga_{11} \Ga_{12} \Ga_{13} \Ga_{14}
= \si_3 \otimes 1
$$
From these Clifford basis vector matrices we can construct the representative $\mathbb{R}(256)$ matrices of the one-hundred-twenty bivector basis generators, $\Ga'_{xy}$, of $spin(12,4)$,
\beq
\begin{array}{rccccccrcccl}
\Ga'_{ij} \!\!&\!\! = \!\!&\!\! \!\! \!\!&\!\! 1 \!\!\!&\!\! \otimes \!\!\!&\!\! \Ga_{ij} \!\!&\!\! = 
\lb
\begin{array}{cc}
\Ga'^+_{ij} & 0 \cr
0 & \Ga'^-_{ij} \cr
\end{array}
\rb_{\vp{\big(}}
\;\;\;\;\; & \Ga'^+_{ij} \!\!&\!\! = \!\!&\!\! \Ga_{ij} \!\!&\!\! = \!\!&\!\!
\lb
\begin{array}{cc}
\Ga^+_{ij} & 0 \cr
0 & \Ga^-_{ij} \cr
\end{array}
\rb_{\vp{\big(}} \cr
\Ga'_{i \, 15} \!\!&\!\! = \!\!&\!\! \!\! \!\!&\!\! 1 \!\!\!&\!\! \otimes \!\!\!&\!\! \Ga_{i} \Ga \!\!& \;\;\;\;\;\;\;\; & \Ga'^+_{i \, 15} \!\!&\!\! = \!\!&\!\! \Ga_{i} \Ga \!\!&\!\! = \!\!&\!\!
\lb
\begin{array}{cc}
0 & -\Ga^+_i \cr
\Ga^-_i & 0 \cr
\end{array}
\rb_{\vp{\big(}} \cr
\Ga'_{i \, 16} \!\!&\!\! = \!\!&\!\! \!\! \!\!&\!\! \si_3 \!\!\!&\!\! \otimes \!\!\!&\!\! \Ga_{i} \!\!& \;\;\;\;\;\;\;\; & \Ga'^+_{i \, 16} \!\!&\!\! = \!\!&\!\! \Ga_i \!\!&\!\! = \!\!&\!\!
\lb
\begin{array}{cc}
0 & \Ga^+_i \cr
\Ga^-_i & 0 \cr
\end{array}
\rb_{\vp{\big(}} \cr
\Ga'_{15 \, 16} \!\!&\!\! = \!\!&\!\! \!\! \!\!&\!\! \si_3 \!\!\!&\!\! \otimes \!\!\!&\!\! \Ga \!\!& \;\;\;\;\;\;\;\; & \Ga'^+_{15 \, 16} \!\!&\!\! = \!\!&\!\! \Ga \!\!&\!\! = \!\!&\!\!
\lb
\begin{array}{cc}
1 & 0 \cr
0 & -1 \cr
\end{array}
\rb \cr
\end{array}
\label{spin124}
\eeq
in which appear the $\mathbb{R}(128)$ $\Ga_{ij}$ and $\Ga_{i}$ matrices from (\ref{bGa}) and (\ref{Ga}). The positive chiral parts of these generators, $\Ga'^+_{xy}$, span the $\mathbb{R}(128)$ positive chiral matrix representation of $spin(12,4)$. They act on a real positive chiral spinor representation space, $\ps = \ps^\ph Q'_\ph \in 128^\mathbb{R}_{S+}$, spanned by the canonical set of unit column matrices, $Q'_\ph$.

The quaternionic real form of the largest simple exceptional Lie algebra, \linebreak $E_{8(-24)}$, is spanned by the one-hundred-twenty independent basis generators, $\Ga'_{xy}$, of $spin(12,4)$ and the one-hundred-twenty-eight $Q'_\ph$, now interpreted as basis generators. The complete set of $E_{8(-24)}$ Lie brackets between these basis generators is:
\beq
\renewcommand{\arraystretch}{1.4}
\begin{array}{rcl}
[\Ga'_{wx}, \Ga'_{yz}] \!\!&\!\! = \!\!&\!\!  2 \eta_{xy} \Ga'_{wz} - 2 \eta_{xz} \Ga'_{wy} + 2 \eta_{wz} \Ga'_{xy} - 2 \eta_{wy} \Ga'_{xz}  \cr
[\Ga'_{xy}, Q'_\ph] \!\!&\!\! = \!\!&\!\! - [Q'_\ph , \Ga'_{xy}] = \Ga'^+_{xy} Q'_\ph = Q'_\ps (\Ga'^+_{xy})^\ps_{\p{\ps} \ph} \cr
[Q'_\ph, Q'_\ps] \!\!&\!\! = \!\!&\!\! - \Ga_{xy} (\Ga'^{+xy})_{\ps \ph} =  - \Ga_{xy} \eta^{xw} \eta^{yz} (\Ga'^+_{wz})^\la_{\p{\la} \ph} g_{\la \ps} \cr
\end{array}
\label{e8}
\eeq
in which $\eta_{xy}$ is a Minkowski metric of signature $(12,4)$ and $g_{\la \ps}$ is part of the $E_{8(-24)}$ Killing form.

The Killing form is a symmetric bilinear operator -- a metric on the $E_{8(-24)}$ Lie algebra as a vector space. Its components between the one-hundred-twenty independent bivector basis elements are
$$
g_{wx \, yz} = (\Ga'_{wx}, \Ga'_{yz}) = \eta_{xy} \eta_{wz} - \eta_{wy} \eta_{xz} 
$$
with signature $(48,72)$ -- the Killing form of $spin(12,4)$. Between bivector and spinor generators the Killing form is zero, $g_{xy \, \ph} = (\Ga'_{xy}, Q'_\ph) = 0$. The $\mathbb{R}(128)$ matrix of Killing form components between spinor generators, $g_{\la \ps} = (Q'_\la, Q'_\ps)$, appearing in (\ref{e8}), is the product of the odd signature Clifford vector matrices,
\beq
g = (\Ga'_1 \Ga'_2 \Ga'_3 \Ga'_{16})^+
=  \Ga'^+_{1 \, 16} \Ga'^+_{2 \, 16} \Ga'^+_{3 \, 16}
= - \si_1 \otimes 1 \otimes 1 \otimes \si_2 \otimes \si_2 \otimes 1 \otimes 1
\label{kill}
\eeq
having signature $(64,64)$. Together, the signature of the $E_{8(-24)}$ Killing form is $(112,136)$ -- consistent with the $112-136 = -24$ label, and with the embedding of $E_7$ in $E_{8(-24)}$.\cite{Adams}

The main observation of this paper is that the GraviGUT algebra, specified by (\ref{s113}) and (\ref{s113s}), is a subalgebra of $E_{8(-24)}$, specified by (\ref{e8}). Specifically, the GraviGUT subalgebra of $E_{8(-24)}$ is spanned by the ninety-one $\Ga'_{ij}$ of $spin(11,3) \subset spin(12,4)$ and sixty-four $Q'_\io$ of $64^\mathbb{R}_{S+} \subset 128^\mathbb{R}_{S+}$, with the subalgebra
\beq
\renewcommand{\arraystretch}{1.4}
\begin{array}{rcl}
[\Ga'_{ij}, \Ga'_{kl}] \!\!&\!\! = \!\!&\!\! 2 \eta_{jk} \Ga'_{il} - 2 \eta_{jl} \Ga'_{il} + 2 \eta_{il} \Ga'_{jk} - 2 \eta_{ik} \Ga'_{jl}  \cr
[\Ga'_{ij}, Q'_\io] \!\!&\!\! = \!\!&\!\! \Ga'^+_{ij} Q'_\io = Q'_\ka (\Ga^+_{ij})^\ka_{\p{\ka} \io} \cr
\end{array}
\eeq
from (\ref{e8}) matching that of the $spin(11,3)$ GraviGUT, having used (\ref{spin124}) to see that $(\Ga'^+_{ij})^\ka_{\p{\ka} \io}=(\Ga^+_{ij})^\ka_{\p{\ka} \io}$. In this completely direct and explicit way, we see that the $spin(11,3)$ GraviGUT, and hence the algebra of gravity and the Standard Model, is inside $E_{8(-24)}$. This embedding of the Standard Model algebra in $E_{8(-24)}$ includes not only the gravitational and Standard Model Lie algebra, but also its correct action on a generation of fermions -- the structure of $E_{8(-24)}$ explains the existence and structure of the spinor, electroweak, and strong fermion multiplets of the Standard Model algebra.

If the complete structure of $E_{8(-24)}$ is interpreted directly, with each of two-hundred-forty-eight generators corresponding to a different kind of elementary particle, then it predicts the existence of many new particles in addition to those of the Standard Model. Of the two-hundred-forty-eight, one-hundred-twenty are basis generators of $spin(12,4)$ and one-hundred-twenty-eight of $128^\mathbb{R}_{S+}$. Of the $spin(12,4)$ generators, ninety-one are for $spin(11,3)$, consistent with a GraviGUT. That Gravi\-GUT has six generators for gravitational $spin(1,3)$, forty-five for $spin(10)$, and forty for a frame-Higgs, including a Higgs 10 multiplet. The $spin(10)$ GUT includes the twelve $g$, $W$, and $B$ bosons of the Standard Model, as well as predicting the existence of three new $W'^{\pm}$ and $Z'$ particles related to $su(2)_R$, and a collection of thirty colored $X$ bosons. The remaining generators of $spin(12,4)$, not in $spin(11,3)$, correspond to twenty more $X$ bosons, one Peccei-Quinn $w$ boson of $spin(1,1)$, and eight for more frame-Higgs' that includes two axions. Of the one-hundred-twenty-eight spinor generators, sixty-four are those of one generation of Standard Model fermions, while the other sixty-four are those of mirror fermions, with opposite charges -- predicted to exist if the structure of $E_{8(-24)}$ is interpreted directly.

Given this explicit embedding of gravity and the Standard Model inside $E_{8(-24)}$, one might wonder how to interpret the paper ``There is no `Theory of Everything' inside E8.''\cite{Distler} In their work, Distler and Garibaldi prove that, using a direct decomposition of $E_8$, when one embeds gravity and the Standard Model in $E_8$, there are also mirror fermions. They then claim this prediction of mirror fermions (the existence of ``non-chiral matter'') makes E8 Theory unviable. However, since there is currently no good explanation for why any fermions have the masses they do, it is overly presumptuous to proclaim the failure of $E_8$ unification -- since the detailed mechanism behind particle masses is unknown, and mirror fermions with large masses could exist in nature. Nevertheless, it was helpful of Distler and Garibaldi to emphasize the difficulty of describing the three generations of fermions, which remains an open problem.

Although it is possible to define a map, based on triality, between the generation of sixty-four Standard Model fermion generators in $E_{8(-24)}$, the sixty-four mirror fermion generators, and sixty-four non-Standard Model boson generators, these cannot be interpreted as three generations of fermions under a direct decomposition of $E_{8(-24)}$. The proposal that these three blocks of generators might correspond to the three generations of fermions, suggested in \cite{E8}, remains a vague hint towards some more mysterious structure, and not a direct identification. The explanation for the existence of three generations of fermions, all with the same apparent algebraic structure, remains largely a mystery. Nevertheless, it is intriguing that the algebraic elements of all gravitational and Standard Model bosons acting correctly on a generation of fermions are found in $E_8$.

\subsection{Superconnection}

The fact that generators corresponding to bosons and fermions coexist in the $E_{8(-24)}$ Lie algebra suggests an interesting geometric construction. When one describes the geometry of a principal bundle over spacetime, the connection is a 1-form valued in the Lie algebra. However, fermion fields are not conventional 1-forms over spacetime, but are fields of anti-commuting numbers -- called Grassmann numbers by physicists. These fields also arise in the BRST approach to gauge theory,\cite{vHolten} in which these anti-commuting fields are sometimes interpreted as 1-forms over the space of connections. As in BRST, it is useful to consider a unified field, called a {\it superconnection}, comprised of the formal sum of connection 1-forms and anti-commuting fields, both valued in the same algebra,
\beq
\udf{A} = \f{H} + \ud{\ps}
\label{sc}
\eeq
Computing the curvature of this superconnection,
\beq
\udff{F} = \f{d} \udf{A} + \ha [ \udf{A}, \udf{A} ] = \ff{F} + \f{D} \ud{\ps} + \ud{\ps} \ud{\ps}
\label{Fs}
\eeq
we find a gauge field curvature, a covariant derivative term, and a term quadratic in $\ud{\ps}$. If the gauge field, $\f{H}$, is the unified bosonic connection, (\ref{ubc}), and the anti-commuting field, $\ud{\ps}$, is a Standard Model fermion multiplet or the $E_{8(-24)}$ subspace corresponding to a fermion multiplet, then the covariant derivative term, $\f{D} \ud{\ps}$, is identical to the covariant Dirac derivative in (\ref{dirac}). The superconnection, (\ref{sc}), constitutes a unified field comprised of all known types of elementary particles, and its curvature, (\ref{Fs}), can be used to describe their dynamics.

\section{Discussion and Conclusion}

In this work we have described an explicit embedding of the Standard Model algebra, including gravity, in the $spin(11,3)$ GraviGUT, and its subsequent embedding in $E_{8(-24)}$. If the geometry of our universe is fundamentally that of an $E_{8(-24)}$ principal bundle, this symmetry must be broken, with the frame-Higgs part of the connection attaining a vacuum expectation value, and the structure reducing to that of gravity and the Standard Model. If $E_{8(-24)}$ unification is interpreted directly, it predicts the existence of many new particles, including $W'$, $Z'$, and $X$ bosons, a rich Higgs sector, and the existence of mirror fermions. The charges of these elementary particles -- their eigenvalues with respect to a collection of mutually commuting force generators -- are presented in the Elementary Particle Explorer\footnote{http://deferentialgeometry.org/epe/}, which displays various GUT and GraviGUT weight diagrams and embeddings. It is a distinct possibility that some of these new particles may be detected at the Large Hadron Collider.

Although the explicit embedding of gravity and the Standard Model in $E_{8(-24)}$ described here is incontrovertible, it raises many questions. Since a direct interpretation implies the existence of mirror fermions, which are not known to exist in nature, it will be necessary to understand how these particles obtain large masses or otherwise work out in the theory. Fortunately, the embedding in $E_{8(-24)}$ also predicts the existence of axions and other Higgs scalars, which interact with the mirrors. These scalars could also help explain the existence of the three generations of fermions -- possibly as scalar-fermion composites -- and their masses. Another potential hint towards addressing the generation problem is the embedding of the Standard Model algebra in the split real form, $E_{8(8)}$, via the Pati-Salam GUT, which is important but has not been extensively discussed here. This issue of generations, as well as particle masses, the action for the theory, and ultimately its quantum description, remain open questions.

The explicit embedding of gravity and the Standard Model in the $spin(11,3)$ GraviGUT and in $E_{8(-24)}$ described here provides a solid base from which to develop these ideas further. The explicit real matrix representation of gravity and the Standard Model, (\ref{sm}), the compatible $Cl(11,3)$ matrix representation, (\ref{Ga}), and the explicit description of the structure of $E_{8(-24)}$ in (\ref{e8}), may be especially useful to current and future researchers, whether considering unification in $E_8$ or other algebras. From the solid foundation of these explicit matrix representations and their embeddings, we might get a better perspective on many of the deep open questions remaining.

\section{Acknowledgements}

The author wishes to thank Andrew Golato for constructive feedback. This research was supported by a grant from The Foundational Questions Institute.

\bibliographystyle{amsplain}

\end{document}